\documentclass[aps,pre,twocolumn,amsmath,amssymb,floatfix]{revtex4-1}
\pdfoutput=1
\usepackage{subfig}
\usepackage{bm}
\usepackage{color}
\usepackage{psfrag}
\usepackage{graphicx}
\usepackage{epsfig}
\usepackage{amssymb}
\usepackage{color}
\usepackage{psfrag}

\begin{document}
\date{March 06, 2020}
\author{ Sergei B. Rutkevich}
\title{Scaling in the 
 massive antiferromagnetic XXZ spin-1/2 chain near the isotropic point}
\begin{abstract}
The scaling limit of the Heisenberg XXZ spin chain  at zero magnetic field  is studied  in the  gapped antiferromagnetic phase.
For a spin-chain ring  having $N_x$ sites, the universal Casimir scaling function, which characterises the leading finite-size correction term in the large-$N_x$ expansion of the ground state energy, is calculated by numerical solution  of the nonlinear integral equation of the convolution type. It is shown  that the same scaling function describes the temperature dependence of the free energy
of the infinite XXZ chain at low enough temperatures in the gapped scaling regime. 
  \end{abstract}
\bibliographystyle{unsrt} 
\maketitle
\section{Introduction}
Integrable models of statistical
mechanics and field theory \cite{Bax,Mussardo10} provide us with a very important source of information about the 
thermodynamic and dynamical properties of the magnetically ordered systems. 
Of  particular importance is any progress in solutions of such models in the
scaling region near the continuous phase transition points, since, 
due to the universality of critical fluctuations,  it does not only yield the exact and detailed
information about the model itself but also about the whole universality class it represents. 

In this paper we address the universal finite-size and thermodynamic properties of the 
anisotropic spin-1/2 XXZ chain in the massive
antiferromagnetic phase in the critical region close to the quantum phase transition at the isotropic point.
The  Hamiltonian of the model has the form
\begin{equation}\label{XXZH}
{H}=\frac{J}{2}\sum_{j=1}^{N_x}\!\!\left(\sigma_j^x\sigma_{j+1}^x+\sigma_j^y\sigma_{j+1}^y+
\Delta\,
\sigma_j^z\sigma_{j+1}^z
\right).
\end{equation}
Here the index $j$ enumerates the spin-chain sites, $\sigma_j^\mathfrak{a}$ are the Pauli matrices,  $\mathfrak{a}=x,y,z$, $J>0$ is the 
antiferromagnetic coupling constant, and
$\Delta$ is the anisotropy parameter. The number of sites will be chosen  even,  $N_x=2M$, and  periodic 
boundary  conditions will be implied, $\sigma_{j+N_x}^\mathfrak{a}\sim \sigma_j^\mathfrak{a}$. 
The massive antiferromagnetic phase is realized in this model at $\Delta>1$. 
Following Lukyanov and Terras \cite{LukTer03},  we shall use the  following convenient parametrization,
\begin{equation}\label{parJ}
J=\frac{1}{{a} \pi}, \quad x=a j,\quad 
 \Delta=\cosh \eta, 
\end{equation} 
where $\eta>0$, $a$ denotes the lattice spacing, and  $x$ is the dimensionful spatial coordinate
of the lattice site $j$.  So, the length of the chain  is $L_x=N_x a$.
The Euclidean evolution in this model is described by the operator $U(y)=\exp(-y{H})$.

{
In the thermodynamic limit $N_x\to\infty$, the antiferromagnetic ground state 
of model \eqref{XXZH} is doubly degenerate
at $\Delta>1$. Its  particle sector  is represented  by the kink-like topological excitations,
which interpolate between two antiferromagnetic vacua \cite{Jimbo94}. Since these excitations  carry  spin
$1/2$, they are usually called "spinons". 

For  finite $N_x$, }  the ground state energy $E_{N_x}(\eta,a)$ of the model \eqref{XXZH} can be  represented as,
\begin{equation}\label{EN}
E_{N_x}(\eta,a)= N_x\,{\mathcal{E}}_{b}(\eta,a)+E_{C}(N_x,\eta,a),
\end{equation}
where $N_x\,{\mathcal{E}}_{b}(\eta,a)$ is the bulk term calculated and studied for all $\Delta$ by C. N. Yang and C. P. Yang 
\cite{YY66_1,YY66_2}  by means of the coordinate Bethe Ansatz. The {\it Casimir energy} term $E_{C}(N_x,\eta,a)$
exponentially vanishes  in the thermodynamic limit $N_x\to\infty$ at fixed $\eta$ and $a$. This term 
describing the finite-size correction to the bulk ground-state energy has been extensively
studied  by many authors \cite{Klum_90,PK91,Klum91,De94,deVe85,Dug15} by means of the technique utilising a certain nonlinear integral  equation (NLIE) of the convolution type.
Most attention in these studies has been concentrated on the  $|\Delta|\le1$ gapless Luttinger liquid phase.
For the massive antiferromagnetic phase that takes place at $\Delta>1$, the NLIE was derived by de Vega and Woynarowich \cite{deVe85},
and further studied  by Dugave {\it et al.}   \cite{Dug15}. 

Note that  study of the Casimir energy finite-size correction term $E_{C}(N_x,\eta,a)$  does not have immediate experimental implications, though  it is   important for the theory and crucial for the interpretation of the results of the computer simulation, which are typically performed on  finite-size
systems. In contrast, calculations of the  {\it per}-site free energy
\begin{equation}\label{frEn}
f(J,\eta,T)=-T \lim_{N_x\to\infty} \frac{1}{N_x}\,\ln {\rm Tr} \,e^{-{H}/T}
\end{equation}
of the infinite XXZ chain has a major importance for the experiments, since it yields the specific heat $c(J,\eta,T)=-T\,\partial_T^2 f(J,\eta,T)$
that can be directly measured in quantum quasi-one-dimensional antiferromagnets. The most systematic approach to 
the thermodynamics of the XXZ spin chain is based on the Thermodynamic Bethe Ansatz (TBA) method, which first version was invented in 1969 by 
C.~N.~Yang and C.~P.~Yang \cite{YY69}, who used  it to study the one-dimensional gas  of delta-interacting bosons. Application of the TBA 
for calculation of the thermodynamic quantities in the XXZ spin chain was started in 1971 by Takahashi \cite{Tak71} and Gaudin \cite{Gau71}, and later
continued by many other authors \cite{TaSu72,Koma87,De94,Klum98,Goe18}. Thermodynamics of the more general XYZ spin-chain model
was studied  by means of the TBA in \cite{TaSu72,Tak91,Klum93}. Further references on the TBA method and its applications in the theory 
of  the integrable spin-chain models can be found in 
monographs \cite{Tak09,Zv10}.

The specific heat $c(J,\eta,T)$, apart from the trivial linear dependence on the coupling constant $J$, depends on temperature $T$ and 
on the anisotropy parameter $\eta$. For a given $\eta>0$, the temperature dependence of $c(J,\eta,T)$ can be found by means of the TBA method, 
see Fig. 4a in \cite{Klum93}, where the  plot of the  specific heat $c(1,{\rm arccosh}(3/2),T)$  obtained this way is shown.

At the isotropic point $\eta=0$, the XXZ chain  \eqref{XXZH}  undergoes a continuous quantum phase transition, and the correlation length diverges.
Close to the isotropic point for $0<\eta\ll1$, the correlation length { becomes} much larger than the lattice spacing, 
and { the spin chain arrives at the massive scaling regime. It will be shown later, that
}
 one should expect { in this regime the following} scaling behaviour of the 
specific heat at low enough temperatures $T\ll J$,
\begin{equation}\label{scC}
c (J,\eta,T)=\frac{T}{\pi J}\,X(t),
\end{equation}
where   $X(t)$ is the universal scaling function depending solely on the scaling parameter $t=T/m(J,\eta)$, and $m(J,\eta)$ is the { spinon} mass, 
which is equal to the half of the gap  in the two-{spinon} excitation energy spectrum. The Casimir energy $E_C(N_x,\eta,a)$ should have the 
analogous  universal scaling behaviour at $\eta\ll 1$; see equation \eqref{scEC} below. Surprisingly, the  scaling behaviour of the specific 
heat and Casimir energy in the XXZ spin chain at $0<\eta\ll 1$ has never been studied in literature, and the corresponding 
universal scaling
functions $X(t)$ and $Y(u)$  remained  unknown. The aim of the present paper is to fill this gap.

First, we  modify the nonlinear integral equation derived by Dugave {\it et al.}   \cite{Dug15}  and proceed in it to the scaling limit  in the massive antiferromagnetic phase in order to 
describe the scaling behaviour of the Casimir energy $E_C(N_x,\eta,a)$. 
The scaling limit is 
understood in the usual way,
\begin{eqnarray}\label{tlm}
&& a\to0,\quad \eta\to+0, \quad N_x\to \infty, \\
&& \xi(\eta,a)=Const, \quad L_x\equiv a N_x=Const.\nonumber
\end{eqnarray}
Here  $\xi(\eta,a)$ is the correlation length, 
 which behaves  \cite{Dug15} at small $\eta>0$ as,
\begin{equation}\label{clas}
\xi(\eta,a) =[2\,m(\eta,a)]^{-1}\,\left[1+O(\exp[-\pi^2/ \eta])\right],  
\end{equation}
and the {spinon} mass $m(\eta,a)$ has the following asymptotic behaviour  \cite{McCoy73}
\begin{equation}\label{met}
m(\eta,a)=\frac{4\,\exp[-\pi^2/(2 \eta)]}{a}
\end{equation}
 at $\eta\to+0$.
 
\begin{figure}[htb]
\centering
\includegraphics[width=\linewidth, angle=00]{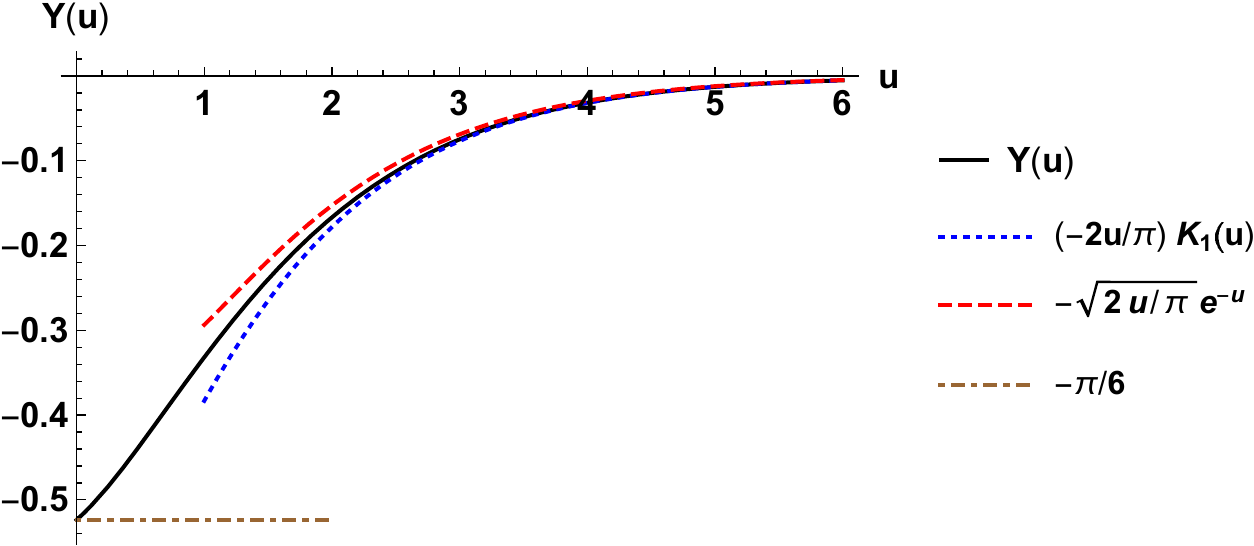}
\caption{\label{scCas} Plot of the Casimir scaling function $Y(u)$ (black solid line).  Its large-$u$ asymptotics
determined by equations \eqref{Yh} and \eqref{Yh1}  are shown by the dotted blue and dashed red  lines, respectively. The value at the critical 
point $Y(0)=-\pi/6$ agrees with the CFT prediction.} 
\end{figure}

It follows from  dimension arguments \cite{Al_Z90} that the Casimir energy takes the scaling form in the limit
\eqref{tlm}, 
\begin{equation}\label{scEC}
E_{C}(N_x,\eta,a)\simeq\frac{Y(u)}{L_x},
\end{equation}
where 
\begin{equation}
u=L_x\,m(\eta,a)= 4 N_x \exp[-\pi^2/(2 \eta)]
\end{equation}
is the scaling parameter and $Y(u)$ is the universal Casimir  scaling function.
We calculated this function numerically by iterative solution of the NLIE written in the scaling limit \eqref{tlm}.
The plot of the resulting Casimir scaling function is shown in Figure \ref{scCas}.

The scaling limit  \eqref{tlm} of model \eqref{XXZH} can be described by 
{ the sine-Gordon quantum field theory {\cite{Lut76,LukTer03}},  
in which the coupling constant $\beta_s^2$ approaches  its upper boundary value $\beta_s^2\to8\pi$.
This }Euclidean  quantum field theory (EQFT) 
 lives on the 
torus having the periods $L_x, L_y$, in the limit $L_y\to \infty$.  Under the choice \eqref{parJ} 
of the coupling constant $J$, the dispersion law of the elementary excitations  in this continuous EQFT 
takes the relativistic form $\omega(p)=\sqrt{p^2+m(\eta,a)^2}$, indicating the rotational symmetry of the 
theory in the $\langle x,y\rangle$ plane. As it was explained by Al. B. Zamolodchikov \cite{Al_Z90}, this allows one   to relate the ground state energy
 of the EQFT [determined in our case by equations \eqref{EN} and \eqref{scEC}]  with the free energy of the chain having the infinite length  $L_y\to\infty$ at a nonzero
temperature $T=1/L_x$; { see equation 
(2.8) in \cite{Al_Z90}}. 
As a result, one arrives at the  following  representation 
for the per-site  free energy  \eqref{frEn} in the scaling regime \eqref{tlm}:
\begin{equation}\label{frE}
f(J,\eta,T)={\mathcal{E}}_b(J,\eta)+\frac{T^2}{\pi J}\, Y(u),
\end{equation}
where $u=m(J,\eta)/T$ is the scaling parameter, and 
\begin{equation}\label{mJ}
m(J,\eta)={4\pi J}\exp\left(-\frac{\pi^2}{2\eta}\right)
\end{equation}
is the {spinon} mass. Note that we have changed notations  in  equations \eqref{frE} and \eqref{mJ} 
using the
coupling constant $J$ instead of the lattice spacing $a=(\pi J)^{-1}$  as the
argument of the functions $f,\mathcal{E}_b, m$.
The free energy reduces to the form \eqref{frE} in the scaling regime, which 
is realised at $a T\ll1$ and $a\,m\ll1$. In terms of the original parameters of the XXZ
chain Hamiltonian, these two strong inequalities read
\begin{equation}\label{sccon}
{T}\ll{ J}, \quad\quad \exp\left(-\frac{\pi^2}{2\eta}\right)\ll 1.
\end{equation}

\begin{figure}[htb]
\centering
\includegraphics[width=1.\linewidth, angle=00]{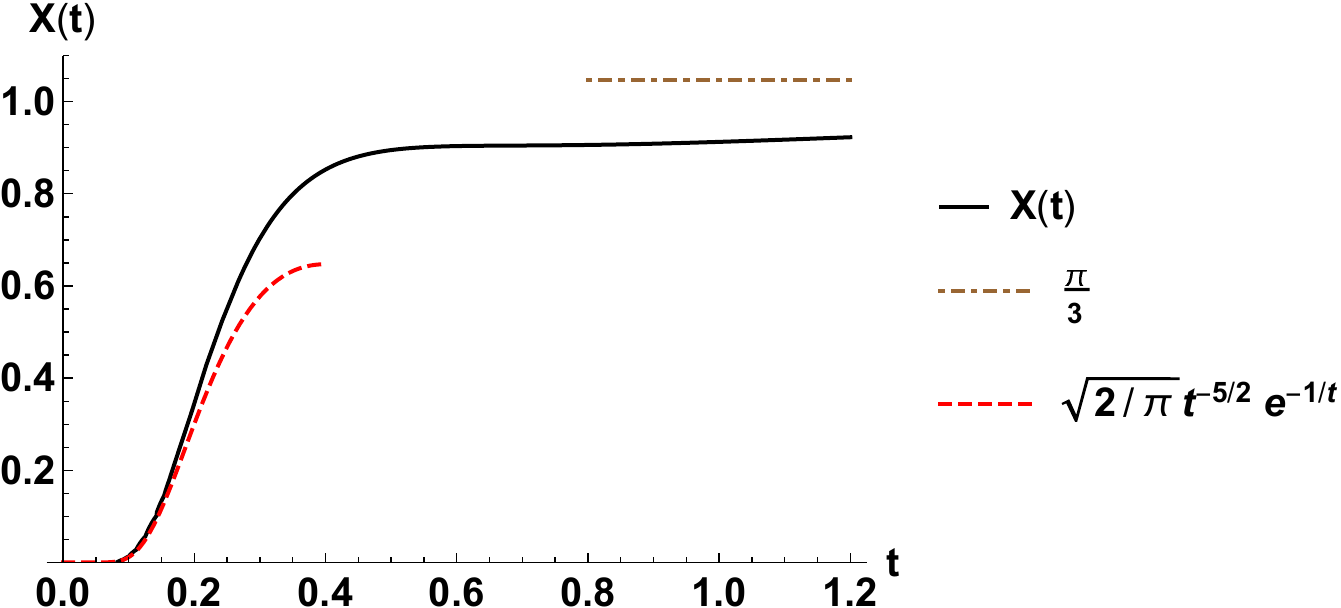}
\caption{\label{CX} Plot of the specific heat scaling function $X(t)$ versus  
the  scaled temperature parameter  $t=T/m(J,\eta)$ (black sold line). The small-$t$ asymptotics \eqref{Xst} of $X(t)$  is shown by 
the red dashed line. 
At large $t$, $X(t)$ approaches very slowly to its CFT value $\pi/3$.} 
\end{figure}

Accordingly, the  specific heat  per  chain site $ c(J,\eta,T)$  must scale 
under conditions \eqref{sccon} to the form \eqref{scC}
where
\begin{equation}\label{XX}
X(t)=\left[- {2Y(u)}+2 u Y'(u)-u^2 Y''(u)\right]\big|_{u=1/t}.
\end{equation}
The plot of the  universal  specific heat scaling function $X(t)$ determined from equation \eqref{XX} is shown in Figure~\ref{CX}.

In what follows, we will describe how these results were obtained and present them in 
 more details.
{
In particular, in section \ref{BAs} we recall briefly a few basic results on the Bethe Ansatz calculation of the ground state 
energy of the finite XXZ spin chain in the gapped antiferromagnetic phase. The well-known representation
of this  ground-state energy  in terms of the solution of a nonlinear integral equation is described 
in section \ref{NIEs}. New results are presented in section \ref{SLs}.  First, we proceed in it to the 
scaling limit in the nonlinear integral equation, and then describe the numerical and 
analytical results for the universal scaling functions characterising the Casimir energy, the  free energy and specific heat of the XXZ spin chain in the antiferromagnetic gapped near-critical regime. 
Section \ref{Con} contains concluding remarks. Finally, in the two Appendixes we describe two alternative 
analytical calculations of the Casimir scaling function $Y(u)$ in the limit  $u\ll1$. 
In  Appendix \ref{Ycal} we exploit to this end the small-$u$ asymptotical analysis of the nonlinear TBA equations, 
while in appendix  \ref{RenGr}  we use the  renormalisation group perturbation-theory technique. 
}
\section{Bethe-Ansatz solution for the ground state at $\Delta>1$\label{BAs}}
The ground state $|\Phi\rangle$ of the $2M$-site chain is characterized by the set of $M$ real Bethe roots $\{\lambda_n\}_{n=1}^M$,
$-\pi/2<\lambda_1<\lambda_2\ldots<\lambda_M<\pi/2$, which solve the equations
\begin{equation}\label{fB}
{\mathcal F}(\lambda_n)+1=0,
\end{equation}
where $n=1,\ldots,M$, and 
\begin{equation}\label{flam}
{\mathcal F}(\lambda)=\left[
\frac{\sin(\lambda-\frac{i \eta}{2})}{\sin(\lambda+\frac{i \eta}{2})}
\right]^{N_x} 
\prod_{j=1}^M \frac{\sin(\lambda-\lambda_j+{i \eta})}{\sin(\lambda-\lambda_j-{i \eta})}.
\end{equation}
Note that $\lambda_n=-\lambda_{M-n+1}$ for the Bethe roots describing the ground state, and 
\begin{eqnarray}
&&{\mathcal F}(\lambda+\pi)={\mathcal F}(\lambda),\;\;
{\mathcal F}(\pi/2)=1,\\
&&{\mathcal F}(\pi-\lambda)=\frac{1}{{\mathcal F}(\lambda)}=\overline{{\mathcal F}(\bar{\lambda})}.\nonumber
\end{eqnarray}

The ground-state energy of the $N_x$-site chain reads
\begin{equation}\label{E}
E_{N_x}=\sum_{n=1}^M \varepsilon_0(\lambda_n),
\end{equation}
where
\begin{equation}
\varepsilon_0(\lambda)=2 J[-\Delta+\cos p_0(\lambda)],
\end{equation}
and
\begin{equation}
\exp[i p_0(\lambda)]=\frac{\sin(\lambda-\frac{i \eta}{2})}{\sin(\lambda+\frac{i \eta}{2})}.
\end{equation}
{Note that 
$
\varepsilon_0(\lambda)=-J\,p_0'(\lambda)\,\sinh \eta.
$}

The counting function $\phi(\lambda)$ can be defined near the real axis by the relations,
\begin{equation}\label{f0}
{\mathcal F}(\lambda)=\exp[2\pi i N_x \phi(\lambda)], \quad \phi(-\pi/2)=0.
\end{equation}
The counting function $\phi(\lambda)$ is analytic in the strip $-\eta/2<{\rm Im}\,\lambda<\eta/2$ 
and quasiperiodic there,
\begin{equation}\label{qph}
 \phi(\lambda+\pi)=\phi(\lambda)+ {\frac{1}{2}}.
\end{equation}
The logarithmic derivative in $\lambda$ of equation \eqref{flam} reads
\begin{equation}\label{phip}
\phi'(\lambda)=\frac{p_0'(\lambda)}{2\pi}-\frac{1}{N_x}\sum_{j=1}^M \mathcal{K}(\lambda-\lambda_j),
\end{equation} 
where
\begin{align}
p_0'(\lambda)=\frac{\cot(\lambda-\frac{i \eta}{2})-\cot(\lambda+\frac{i \eta}{2})}{i},\\
 \mathcal{K}(\lambda)=\frac{\cot(\lambda-{i \eta})-\cot(\lambda+{i \eta})}{2\pi i}.
\end{align} 
\section{Nonlinear integral equation\label{NIEs}}
Assuming that the counting function  $\phi(\lambda)$ corresponding to the ground state strictly increases 
at real $\lambda$, and taking into account \eqref{f0},  \eqref{qph}, one concludes that equation \eqref{fB} has exactly $M$
real solutions in the interval $-\pi/2<\lambda<\pi/2$, and these solutions coincide with the Bethe roots $\{\lambda_n\}_{n=1}^M$.
Application of  Cauchy's integral formula to the sums in the right-hand sides of \eqref{phip} and  \eqref{E} leads  to the following integral representations of these equations   \cite{Dug15}:
\begin{eqnarray}
&&\phi'(\lambda)=\frac{p_0'(\lambda)}{2\pi}-\int_{-\pi}^0 d\mu \,  \mathcal{K}(\lambda-\mu)\phi'(\mu)+\label{ie2}\\\nonumber
&&\frac{1}{\pi  N_x}\int_{-\pi}^{0}
d\mu \,  \mathcal{K}(\lambda-\mu)\, {\rm Im}\,\partial_\mu \ln[1+{\mathcal F}(\mu+i 0)],\\
&&E_{N_x}=N_x \int_{-\pi}^0 d\lambda\,\varepsilon_0(\lambda) \phi'(\lambda)-\label{E1} \\\nonumber
&&\frac{1}{\pi  }\int_{-\pi}^{0}
d\lambda \, \varepsilon_0(\lambda) \, {\rm Im}\, \partial_\lambda \ln[1+{\mathcal F}(\lambda+i 0)].
\end{eqnarray} 
Let us define the linear integral operator $K$ that acts on a $\pi$-periodical function $\psi(\lambda)$ of  $\lambda\in \mathbb{R}$ as follows,
\[
K [\psi](\lambda)=\int_{-\pi}^0d\mu \,  \mathcal{K}(\lambda-\mu)\psi(\mu).
\]
By action with the operator $(1+K)^{-1}$ on  both sides of equation \eqref{ie2}, and subsequent integration in $\lambda$, 
one modifies it to the form
\begin{eqnarray}\label{lnf1}
-i \ln {\mathcal F}(\lambda|\eta,N_x)=2\pi N_x \phi_0(\lambda|\eta)+\\\nonumber
2\int_{-\pi}^{0}
d\mu \, Q(\lambda-\mu|\eta)\, {\rm Im}\, \ln[1+{\mathcal F}(\mu+i 0|\eta,N_x)],
\end{eqnarray}
where 
\begin{align}\label{sQ}
&Q(\lambda|\eta)\equiv(1+K)^{-1} [\mathcal {K}](\lambda)= \sum_{n=-\infty}^\infty \,e^{2 i n \lambda}\,\frac{e^{-2\eta|n|}}{\pi(1+e^{-2\eta|n|})},\\\label{php}
&\phi_0'(\lambda|\eta)\equiv\frac{1}{2\pi}(1+K)^{-1} [p_0'](\lambda)= \sum_{n=-\infty}^\infty  \,\frac{e^{2 i n \lambda}}{2\pi \cosh(\eta\, n)}\nonumber\\
&=\sum_{l=-\infty}^\infty  \,
\frac{1}{2\eta \cosh[\pi(\lambda-\pi l)/\eta]},\\
&\ln {\mathcal F}(-\pi/2|\eta,N_x)=0,\quad \phi_0(-\pi/2|\eta)=0. \label{BC}
\end{align}

Similarly, the ground-state energy \eqref{E1} can be  represented in the form \eqref{EN},
where
\begin{equation}\label{Ome}
\mathcal{E}_{b}(\eta,a)= \int_{-\pi}^0 d\lambda\,\varepsilon_0(\lambda) \phi_0'(\lambda),
\end{equation}
is the ground state energy per site in the infinite chain, 
and the finite-size correction (Casimir energy) reads
\begin{align} \label{Cas}
E_{C}(N_x,\eta,a)=  -{2} J \sinh\eta \int_{-\pi}^{0} d\lambda\,\, \phi_0''(\lambda)\cdot\\\nonumber
\,{\rm Im}\,\ln\left[1+{f}(\lambda{+i 0})\right].
\end{align}
\section{Scaling limit \label{SLs}}
In the scaling limit \eqref{tlm}, the solution ${\mathcal F}(\lambda|\eta,N_x)$ of equation \eqref{lnf1} approaches   very fast to its
 bulk limit $\exp[2 \pi i N_x\,\phi_0(\lambda|\eta)]$ everywhere in the real $\lambda$ axis, apart from the small vicinities of the points
 $\lambda^{(n)}=-\pi/2+ \pi n$. To describe  the scaling limit of equation  \eqref{lnf1} near one of such points  $\lambda^{(0)}=-\pi/2$,
 let us 
make in it the linear change of the rapidity variables $\lambda,\mu$, 
\begin{equation}\label{cha}
\lambda=-\frac{\pi}{2}-\frac{\eta\, \alpha}{\pi}, \quad \mu=-\frac{\pi}{2}-\frac{\eta\, \alpha'}{\pi},
\end{equation}
where $\alpha,\alpha'$ are the rescaled rapidities.

The  function ${\mathcal F}(\lambda|\eta,N_x)$ reduces 
in the vicinity of  the point $\lambda^{(0)}$ in the scaling limit \eqref{tlm} to the form
\begin{equation}
{\mathcal F}(\lambda|\eta,N_x) _{\lambda=-\pi/2-\eta\,\alpha/\pi}=\frac{1}{ \mathfrak{f}(\alpha|u)}+({\rm corrections\,to\, scaling}).
\end{equation}

{
The scaling limit of the first term on the right-hand side of the integral equation \eqref{lnf1} reads,
\begin{equation}\label{scfi}
2\pi N_x \,\phi_0(\lambda|\eta)_{\lambda=-\pi/2-\eta\,\alpha/\pi}=-u  \sinh \alpha\,[1+O(ma)^2].
\end{equation}
To prove this, let us  note that 
 the leading contribution to the sum in the second line of \eqref{php} comes at
 $\eta\to+0$ and $\lambda\approx-\pi/2$
 from the two terms with $l=-1,0$. Then simple calculations yield,
\begin{equation}
\phi_0'(\lambda|\eta)_{\lambda=-\pi/2-\eta\,\alpha/\pi}=\frac{m a}{2\eta}  \cosh\alpha \,\,[1+O(am)^2].
\end{equation}
Integration  of this equality with respect to  $\lambda$ with \eqref{BC} taken into account leads 
to \eqref{scfi}.

In order to find the scaling limit of the kernel $Q(\lambda-\mu|\eta)$ in equation \eqref{lnf1}, we 
replace the sum on the right-hand side of \eqref{sQ} by the integral,
\[
Q(\lambda-\mu|\eta)=\frac{\pi}{\eta}\, {\mathcal Q}(\alpha-\alpha')+O(\eta),
\]
where $\lambda,\mu$  according to \eqref{cha}  are related to $\alpha,\alpha'$, and
\begin{equation}\label{ps}
{\mathcal{Q}}(\Lambda)=\frac{1}{\pi}\int_{0}^\infty dy\,
\cos(2 y\Lambda)\,\frac{e^{-\pi y}}{\cosh(\pi  y)}.
\end{equation}
The integral on the right-hand side can be explicitly calculated,
\begin{equation}\label{psa}
{\mathcal{Q}}(\Lambda)=\lim_{\gamma_s\to\infty}
\frac{1}{2 \pi i}\frac{\partial \ln S(\Lambda,\gamma_{s})}{\partial\Lambda},
\end{equation}
where $S(\Lambda, \gamma_s)$ is the soliton-soliton scattering amplitude in the 
sine-Gordon model \cite{Zam77,Sm92},   
\begin{align}\label{Sc}
&S(\Lambda, \gamma_s)=-\exp
\left[
-i\int_0^\infty \frac{dy}{ y} \frac{\sin(2 \Lambda y)\sinh[(\pi-\frac	{\gamma_s}{8})y]}{\cosh (\pi y)\sinh(\gamma_s y/8)}\right], \\
&\lim_{\gamma_s\to\infty}S(\Lambda, \gamma_s)=-\frac{\Gamma\left(1+\frac{i \Lambda}{2\pi}\right)
\Gamma\left(\frac{1}{2}-\frac{i \Lambda}{2\pi}\right)}{\Gamma\left(1-\frac{i \Lambda}{2\pi}\right)
\Gamma\left(\frac{1}{2}+\frac{i \Lambda}{2\pi}\right)}.\label{lim}
\end{align}
Here the parameter $\gamma_s$  is simply related to the  coupling constant $\beta_s^2$ in the 
sine-Gordon model  $\gamma_s=\beta_s^2/(1-\frac{\beta_s^2}{8\pi})$. It is well known \cite{KiRe87,Dio98}, that equation \eqref{Sc}
describes also the amplitude of the spinon-spinon scattering in the XXZ model \eqref{XXZH} in the 
gapless phase $|\Delta|<1$, if the parameter $\gamma={\rm arccos}\,\Delta$ is chosen so that
\begin{equation}\label{gam}
 \gamma=\frac{8\pi^2}{8\pi+\gamma_s}=\pi\left(1-\frac{\beta_s^2}{8\pi}\right).
\end{equation}
The limit $\gamma_s=\infty$ corresponds to the isotropic antiferromagnetic point  of the XXZ spin chain, in which 
$\gamma=0$ and ${\Delta=1}$.  The spinon-spinon
scattering phase factor \eqref{lim}  at this point  of model \eqref{XXZH} was first obtained by Faddeev and Takhtajan \cite{FT81}.

The nonlinear integral equation \eqref{lnf1} reduces in the scaling limit to the form
}
\begin{eqnarray}\label{lnf2}
&&-i \ln {\mathfrak{f}}(\alpha|u)=u \sinh \alpha+\\\nonumber
&&2\int_{-\infty}^{\infty}
d\alpha' \, \mathcal{Q}(\alpha-\alpha') \,{\rm Im}\,\ln[1+\mathfrak{f}(\alpha'+i 0|u)],
\end{eqnarray}
with real $\alpha,\alpha'\in \mathbb{R}$, and the scaling limit of the Casimir energy \eqref{Cas} reads
\begin{eqnarray}\label{Cas1}
E_{C}(N_x,\eta,a)= -\frac{m}{\pi }\int_{-\infty}^{\infty} d\alpha \, \sinh \alpha \cdot \\
{\rm Im}\ln\left[1+\mathfrak{f}(\alpha+i 0|u)\right] \nonumber.
\end{eqnarray}

Equations (\ref{lnf2})  and \eqref{Cas1} coincide with the ${\gamma\to 0}$ limit of 
equations (5.9) and (5.8) 
obtained by Destri  and de Vega \cite{De94} for the massive Thirring  (sine-Gordon) model. 
However, concentrating in their article  on the massive Thirring model { with a finite $\gamma>0$}  and on the gapless 
case of the XXZ spin chain, the authors of \cite{De94} did not apply their results to describe 
the massive scaling regime of the XXZ chain, which we address here.
  Perhaps, for this reason, Destri  and de Vega
did not study in \cite{De94}  the non-trivial $\gamma\to0$ limit of 
their integral equation (5.9), which is relevant to the XXZ model in the massive scaling regime.

Let us analytically continue the function ${\mathfrak{f}}(\alpha|u)$  into the strip 
$|{\rm Im}\, \alpha|\le \pi/2$ and introduce  two auxiliary complex functions (the pseudoenergies)
$\varepsilon(\beta|u), \bar{\varepsilon}(\beta|u)$,
\begin{equation}
\varepsilon(\beta|u)=-\ln\mathfrak{f}(\beta+i \pi/2|u),\quad
\bar{\varepsilon}(\beta|u)=\ln \mathfrak{f}(\beta-i \pi/2|u).
\end{equation}
At real $\beta$, these functions are complex conjugate to one another.
They must satisfy the system of two nonlinear integral TBA equations
(compare with equation (3.3) and  (3.4) in \cite{Al_Z90}),
which follow from \eqref{lnf2},
\begin{subequations}\label{tBa}
\begin{align}\nonumber
& \varepsilon(\beta|u)=u \cosh \beta
- \int_{-\infty}^{\infty}
d\beta' \, \mathcal{Q}(\beta-\beta') \ln[1+e^{-\varepsilon(\beta'|u)}]\\
 &+\int_{-\infty}^{\infty}d\beta' \, \mathcal{Q}(\beta-\beta'+i \pi - i 0)  \ln\large[1+e^{-\bar{\varepsilon}(\beta'|u)}\large],\label{tBA1}\\\nonumber
&\bar{\varepsilon}(\beta|u)=u \cosh \beta
- \int_{-\infty}^{\infty}
d\beta' \, \mathcal{Q}(\beta-\beta') \ln\large[1+e^{-\bar{\varepsilon}(\beta'|u)}\large]\\
&+ \int_{-\infty}^{\infty}d\beta' \, \mathcal{Q}(\beta-\beta'-i \pi + i 0)  \ln\large[1+e^{-{\varepsilon}(\beta'|u)}\large].\label{tBA2}
\end{align}
\end{subequations}
In turn, the  Casimir energy  \eqref{Cas1} takes the   form \eqref{scEC}, with 
 the  scaling function
\begin{eqnarray}\label{Ysc}
Y(u) =
- \frac{u}{\pi} \int_{-\infty}^\infty {d\beta}\,\cosh\beta\,\,{\rm Re}\,  \ln[1+e^{-\varepsilon(\beta|u)}].
\end{eqnarray}

The nonlinear integral equations  (\ref{tBa}) with a different first term in the 
right-hand sides, however,  were  studied by Kl\"umper \cite{Klum93,Klum98}, who used them to calculate the temperature dependence
of the specific heat and magnetic susceptibility in the isotropic antiferromagnetic XXX spin-1/2 chain. 

Note that the nonlinear integral equations \eqref{tBa}
can be derived in a completely different way exploiting   Kl\"umper's results \cite{Klum93} for the general XYZ spin chain.
 If one starts from the TBA equations (3.19) derived in \cite{Klum93} for the XYZ chain, proceeds to the limit corresponding 
 to the  XXZ chain in the gapped antiferromagnetic phase $\Delta>1$, and afterwards proceeds to the scaling limit $\eta\to +0$, one arrives at equations
  \eqref{tBa}. In turn, formula (3.21) in \cite{Klum93} representing the free energy of the XYZ chain reduces 
  after these two limiting procedures to the scaling form \eqref{frE} with the scaling function
  $Y(u)$ given by  \eqref{Ysc}.

The system of nonlinear integral equations  \eqref{tBa} can be solved numerically by iterations.
The convergence of iterations is perfect at large and intermediate values of the scaling parameter $u$, but retards at very small $u$.
The plot of the resulting Casimir scaling function $Y(u)$ is shown in Figure \ref{scCas}. 

The scaling  function $Y(u)$  exponentially decays at large $u\to+\infty$, 
\begin{eqnarray}\label{Yh}
Y(u)=-\frac{2u}{\pi} K_1(u)+O(e^{-2u})=\\
-\sqrt{\frac{2u}{\pi}} \,e^{-u}\,[1+O(1/u)],\label{Yh1}
\end{eqnarray}
where $K_1(u)$ is the Macdonald function.  As in the case of the  massive Thirring model  at a finite $\gamma>0$ 
(see equations (6.9)-(6.11) in \cite{De94}),
the large-$u$ asymptotics \eqref{Yh} can be easily obtained by replacing 
the function $\varepsilon(\beta|u)$ in  \eqref{Ysc} by its the "zeros 
iteration" $u \cosh \beta$   and  then expanding the resulting logarithm in the integrand to the first order, 
$\ln[1+\exp(-u \cosh \beta)]\to \exp(-u \cosh \beta)$. 

At the isotropic point $u=0$, the scaling function takes the value 
\begin{equation}\label{pi6}
Y(0)=-\pi/6,
\end{equation} 
in agreement with the CFT prediction \cite{Al_Z90,De94} for the Gaussian field theory with the central charge $c=1$.
 { A rather involved} perturbative calculation of { three} further terms in the small-$u$ expansion of $Y(u)$  { is described in Appendix \ref{Ycal}. The final results read,
}
\begin{equation}
Y(u)=-\frac{\pi}{6}+ \frac{\pi}{16\,  R(u)^3}+
\frac{3\pi \,\ln[2 R(u)] }{32 \,R(u)^4}+ 
\frac{a_4}{R(u)^4}+\ldots,\label{Y0}
\end{equation}
where   $a_4\approx-0.193$, and 
$
R(u)= \ln(2/u). 
$

\begin{figure}
\centering
\subfloat[]
{\label{usm}
\includegraphics[width=1.\linewidth]{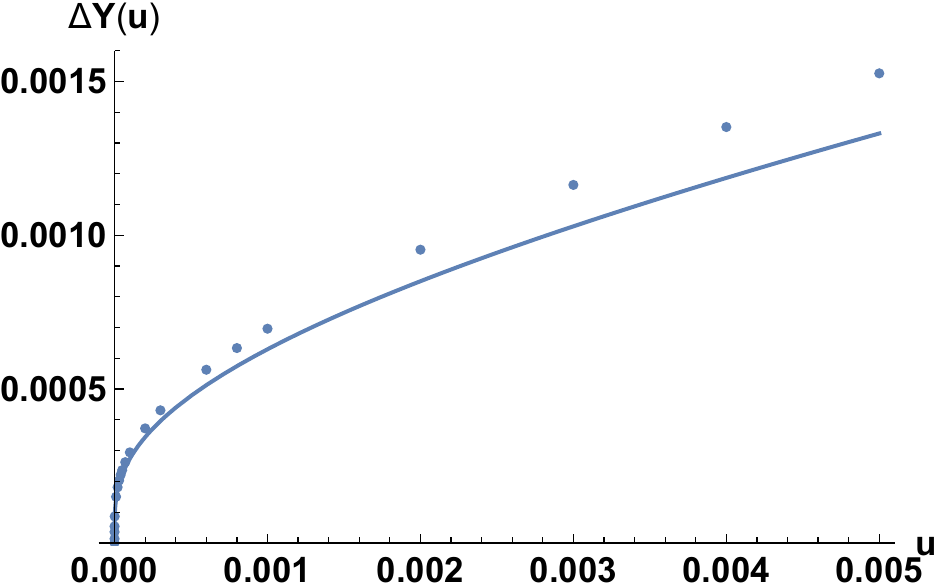}}

\subfloat[]{
	\label{ddY}
\includegraphics[width=1.\linewidth]{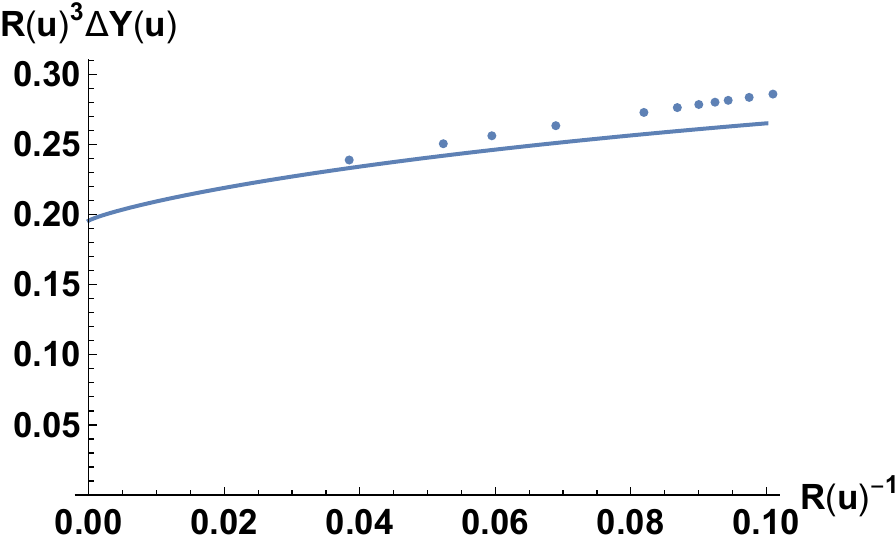}}

\caption{Comparison of analytical and numerical results for 
the scaling function  $\Delta Y(u)=Y(u)+\pi/6$ 
at small $u$. \\
(a) Numerical data (dots)  for the 
function $\Delta Y(u)$ and its asymptotical behaviour  (\ref{Y0}) (solid curve) at $u<0.005$.\\
(b) Numerical data (dots) for the  function $R(u)^3\,\Delta Y(u)$ plotted against 
$R(u)^{-1}$ at $10^{-11}\le u\le 10^{-4}$. The solid curve displays the small-$u$ asymptotic formula  (\ref{RdY}) 
for this function.}
\label{3}
\end{figure}

The logarithmic singularity of the Casimir scaling function at $u=0$  predicted by 
equation \eqref{Y0} is too weak to  be resolved in Figure \ref{scCas}. 
However, this singularity is clearly seen in { Figure \ref{usm}, which displays  at small $u<0.005$ the deviation of 
$Y(u)$ from its  CFT limit \eqref{pi6},  $\Delta Y(u)=Y(u)+\pi/6$.}
The numerical data shown by dots slowly approach with decreasing $u$  the solid curve 
representing the asymptotical formula  (\ref{Y0}). { The same tendency remains at very small 
$u$, as one can see in 
Figure \ref{ddY}.  The 
dots in this Figure display the numerical data for the function 
$R(u)^3\, \Delta Y(u)$  in the interval $10^{-11}\le u\le 10^{-4}$ plotted against $R(u)^{-1}$.
The solid curve represents the small-$u$ asymptotics for this function, 
\begin{equation}\label{RdY}
R(u)^3\, \Delta Y(u)\approx \frac{\pi}{16}+ \frac{3\pi}{32} \,R(u)^{-1}  \ln[2 R(u)] -0.193\,R(u)^{-1}, 
\end{equation}
 corresponding to \eqref{Y0}. }
The asymptotic { low- and high}-temperature behaviour of the free energy $\Delta f(T)= f(T)-f(0)$ per site can be read from \eqref{frE},
\eqref{Yh}, and \eqref{Y0},
\begin{align}\label{def}
\Delta f(T)=-\frac{T^2}{6 J}\,\bigg[1-\frac{3}{8 \,\ln^3(2T/m)}-\\
\frac{9 \ln[2 \ln(2T/m)]}{16 \,\ln^4(2T/m)} -\frac{6\, a_4}{\pi \,\ln^4(2T/m)}  +\ldots\bigg], \nonumber
\end{align}
at $m\ll T \ll J$, and 
\begin{equation}
\Delta f(T)=-\frac{1}{\pi J }\sqrt{\frac{2 m}{\pi}}\, T^{3/2}\, e^{-m/T}\,[1+O(T/m)], \label{Delf}
\end{equation}
at $ T \ll m\ll J$, where $m$ is given by \eqref{mJ}.  The corresponding small-$t$ asymptotics  of the 
 scaling function $X(t)$ takes the form, 
\begin{equation}\label{Xst}
X(t)=-\sqrt{\frac{2}{\pi}}\,t^{-5/2}e^{-1/t}[1+O(t)] \quad {\rm at }\;\; t\to0, 
\end{equation}
and $X(t)$ slowly approaches  the CFT value $\pi/3$ at very large $t\to\infty$.

For the  total free energy $ F(T,L_y)=N_y\, \Delta f(T)$ of the spin chain of the length $L_y\to \infty$, which has   $N_y=\pi J L_y$ sites,
the low-temperature asymptotics following from \eqref{Delf} reads,
\begin{equation}\label{FE}
F(T,L_y)=-2L_y \sqrt{\frac{ m}{2\pi}}\, T^{3/2}\, e^{-m/T}\,[1+O(T/m)].
\end{equation}
This result has a transparent physical interpretation. One can easily see, that the right-hand side of equation \eqref{FE} is 
just  the grand canonical potential $\Omega(T,L_y)$ of the classical ideal gas
of two kinds of nonrelativistic particles ({spinons} with spins oriented up  and down) having the same mass $m$ and 
the chemical 
potential $\mu=m$,  which move in one dimension in the line of the length $L_y$. 

{
It is interesting to compare two asymptotical formulas for the ground-state energy of the 
 XXZ spin chain of finite length $L_x$ supplemented with  periodic boundary conditions. 

The first one 
\begin{equation}\label{Ei}
E_{N_x}(0,a)= N_x\,{\mathcal{E}}_{b}(0,a)- \frac{\pi  }{6L_x}-\frac{\pi}{16L_x \,\ln^3 \frac{L_x}{a}}+\ldots
\end{equation} 
contains three initial terms
in the  large-$L_x$ 
expansion of the XXZ spin-chain ground-state energy at the isotropic point $\eta=0$. 
The third term on the right-hand side was first obtained by Affleck {\it et al.} \cite{Affleck_1989} in the conformal perturbation theory approach, and later  confirmed (for the analogous  low-temperature expansion of the  free energy) by Kl\"umper \cite{Klum98} 
 in the discrete-lattice 
TBA calculations. 

The second formula
\begin{equation}\label{EA}
E_{N_x}(\eta,a)= N_x\,{\mathcal{E}}_{b}(\eta,a)- \frac{\pi  }{6L_x}-\frac{\pi}{16L_x \,\ln^3\frac{L_x}{\xi(\eta)}}+\ldots
\end{equation} 
holds in the gapped antiferromagnetic phase in the scaling regime $0<\eta\ll 1$ 
at $L_x\ll\xi(\eta)$. Equation \eqref{EA} results from the substitution of  two initial terms of the small-$u$ expansion \eqref{Y0}  into \eqref{EN} and \eqref{scEC}. 

Though formulas \eqref{Ei} and \eqref{EA} look remarkably similar, there are two important differences between them. 
\begin{enumerate}
\item The third term on the right-hand side of  \eqref{Ei} explicitly depending on the lattice 
spacing $a$ describes the discrete-lattice correction to scaling in the ground-state
 energy. In contrast, the third term  on the right-hand side of  \eqref{EA}
 does not depend on $a$ and 
 describes the universal scaling behaviour of the ground-state energy at
 $0<\eta\ll1$.
 \item The ratio $ {L}/{a}$ is the large parameter in equation \eqref{Ei}, and the third term
 on its right-hand side is, therefore,  negative. In contrast, the ratio ${L}/{\xi(\eta)}$ is the  small parameter in the 
 asymptotic expansion   \eqref{EA}, and the analogous correction term in the latter 
 is positive.
\end{enumerate} 
It turns out, that  expansion \eqref{EA} can be also obtained by means of the perturbative CFT
technique applied  in \cite{Affleck_1989} for the derivation of \eqref{Ei}. The difference is that the log-correction
term in \eqref{Ei} was caused by the marginally irrelevant  perturbation of the Gaussian CFT Hamiltonian,
whereas in the case of \eqref{EA} the  perturbing field is  marginally relevant. The field-theoretical
derivation of formula \eqref{EA} is described in Appendix \ref{RenGr}. }
\section{Conclusions\label{Con}}
Considering the Heisenberg
XXZ spin-chain ring in the gapped antiferromagnetic phase $\Delta>1$ close to the quantum phase transition point $\Delta=1$, we expressed its ground-state energy universal  Casimir  scaling function  in terms of the solution of the nonlinear integral equation. We calculated this Casimir scaling function numerically by iterative solution of the 
nonlinear integral equation, and also   analytically determined its asymptotical form
at large and small values of the scaling parameter. Then, using the correspondence  in the scaling regime 
between the ground state energy of the finite ring of length $L_x$ with 
the free energy of the infinite  chain at temperature $T=1/L_x$, we calculated the universal
scaling function $X(t)$ describing the temperature dependence of the specific heat of the infinite chain at low temperatures $T\ll J$ and  
$0<\Delta-1\ll 1$. 

In contrast to many previous studies \cite{Tak71,Gau71,TaSu72,Klum93,De94,Klum98} of  the specific heat in the XXZ spin chain, 
our results   are universal, since we have limited  analysis to  the scaling regime. 
Due to its universality, the obtained scaling function $X(t)$ should describe exactly 
 the specific heat temperature dependences in  those  quasi-one-dimensional 
 magnetic compounds in the scaling regime close to the isotropic point, whose  magnetic Hamiltonian 
  falls into the universality class of the XXZ spin-1/2 chain model.  

In presenting the results, 
we followed the important recommendation of Tracy and McCoy in \cite{TrMcCoy75}: "We strongly recommend that all data be presented in scale-variable and scale-function language". 

It would be interesting to 
experimentally observe  in  quasi-one-dimensional  antiferromagnetic compounds  
 the universal specific-heat scaling temperature dependence \eqref{scC}. 
It would be also interesting and important for the experimental applications to study corrections 
in small $\eta$ to the scaling dependences \eqref{scEC} and  \eqref{scC}. 

\begin{acknowledgments} I am thankful to H. W. Diehl for interesting discussions, and to 
A. Kl\"umper for numerous suggestions leading to improvement of the text.
\end{acknowledgments}
\appendix
\section{ Perturbative  derivation of \eqref{Y0}  \label{Ycal}}
In this appendix we perform the asymptotic analysis of the nonlinear TBA integral equations \eqref{tBa} at 
 a small  ${u\to0}$,
and describe briefly the derivation    of formula \eqref{Y0} for the Casimir scaling function \eqref{Ysc}. Our calculations are based to some extent on the techniques developed by Destri and de Vega \cite{De94} and by Kl\"umper \cite{Klum98} for  different TBA integral equations. We will comment on these works later.

Let us  rewrite the integral equations  \eqref{tBa}  in the equivalent form,
\begin{subequations}\label{tB}
\begin{align}\nonumber
 &\varepsilon(\beta|u)=u \cosh \beta
- 2 \int_{-\infty}^{\infty}
d\beta' \, \mathcal{Q}(\beta-\beta') \,  \label{tB1}
{\rm Re}\,\chi(\beta'|u) 
 -\\
& \int_{-\infty}^{\infty}d\beta' \, g(\beta-\beta' - i 0)\, \overline{\chi(\beta'|u)},\\\nonumber
&  \Bar{\varepsilon}(\beta|u)=u \cosh \beta
- 2 \int_{-\infty}^{\infty}
d\beta' \, \mathcal{Q}(\beta-\beta')\, {\rm Re}\,\chi(\beta'|u)
 -\\
 &\int_{-\infty}^{\infty}d\beta' \, \overline{g(\beta-\beta' - i 0)}\, \chi(\beta'|u),
\end{align}
\end{subequations}
where
\begin{eqnarray}
&&g(\Lambda)=-\frac{1}{2 \Lambda(\Lambda+i \pi)}, \\
&&\chi(\beta|u)=\ln\left[1+e^{-{\varepsilon}(\beta|u)}\right].\label{ch}
\end{eqnarray}

The  integral kernel $\mathcal{Q}(\Lambda)$  determined by equations 
\eqref{psa} and \eqref{lim}  is  real at real $\Lambda$, and  
behaves  at large $|\Lambda|$ as, 
\begin{equation}\label{Qas}
\mathcal{Q}(\Lambda)=\frac{1}{4\Lambda^2}+\frac{\pi^2}{8 \Lambda^4}+O(\Lambda^{-6}).
\end{equation}
The reflection symmetry   of the integral kernels in  \eqref{tB} 
\begin{equation}\label{symk}
\mathcal{Q}(\Lambda)=\mathcal{Q}(-\Lambda), \quad g(\Lambda-i0)=\overline{g(-\Lambda-i0)}
\end{equation}
ensures the reflection symmetry of the pseudoenergy
\begin{equation}\label{syme}
\varepsilon(\beta|u)=\bar{\varepsilon}(-\beta|u)
\end{equation}
at real $\beta$.

\begin{figure}
\centering
\subfloat[]{
	\label{sol}
\includegraphics[width=1.\linewidth]{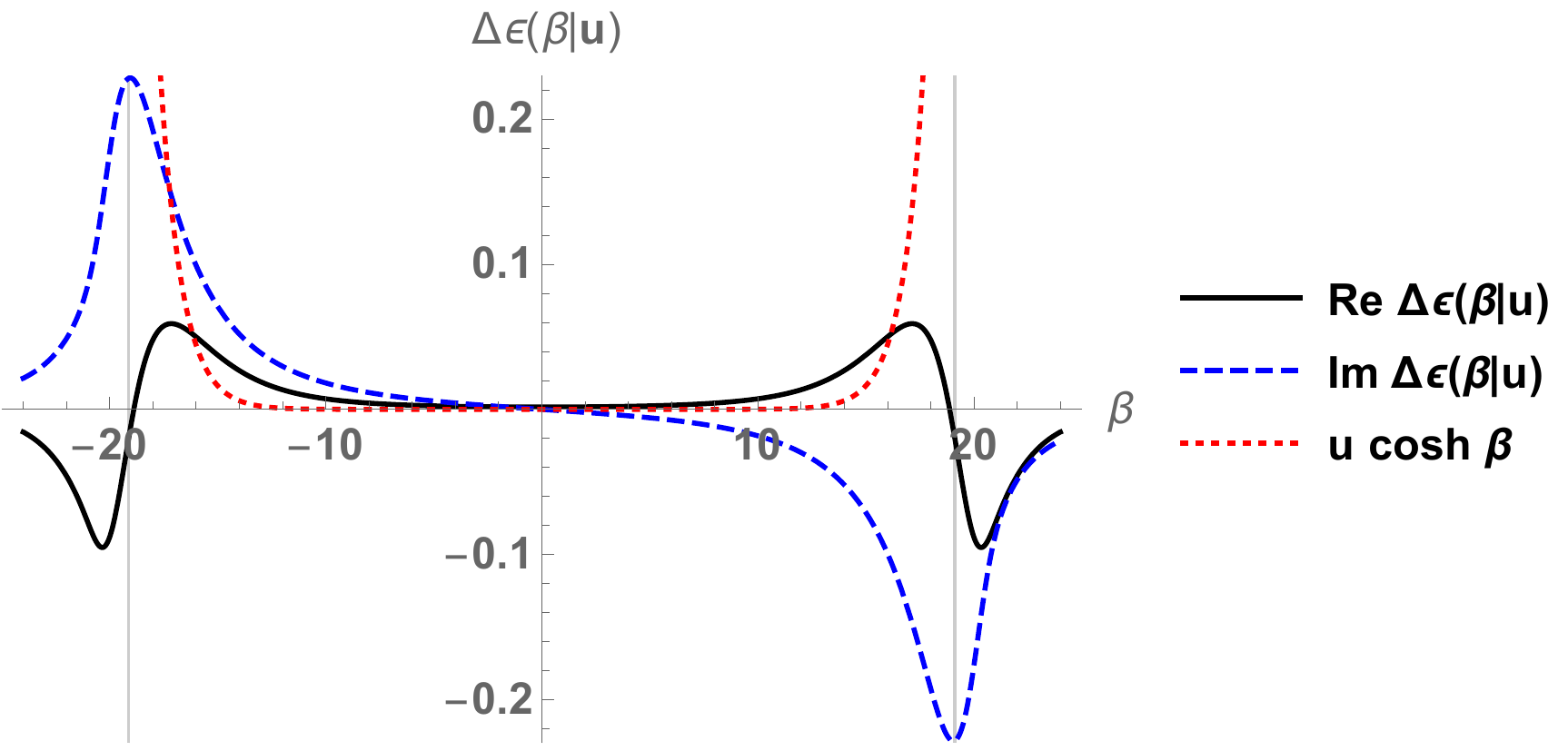}}

\subfloat[]{
	\label{solk}
\includegraphics[width=1.\linewidth]{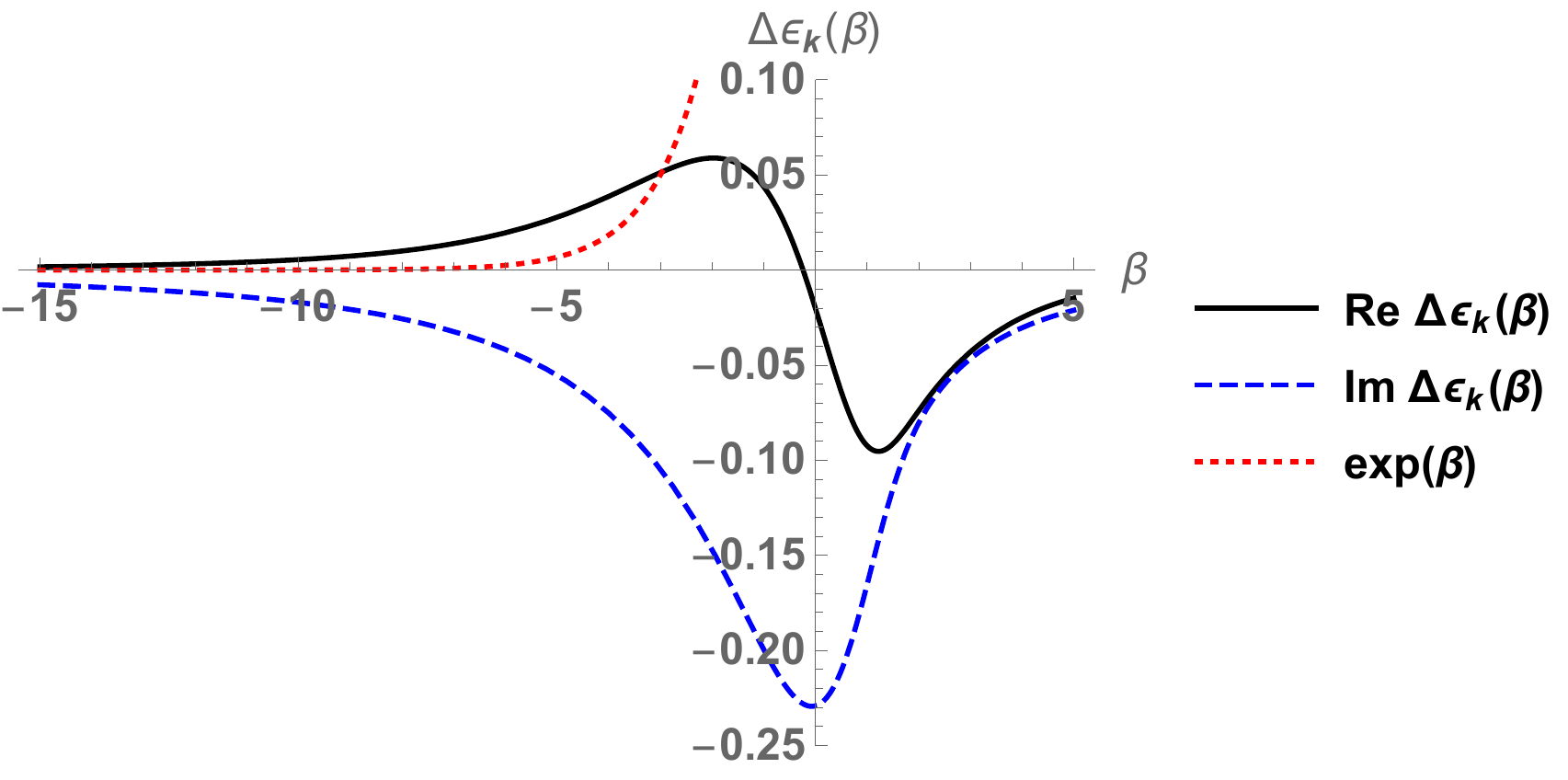}}

\caption{Real and imaginary parts of the pseudoenergy versus rapidity $\beta$ (a) at $u=10^{-8}$ and (b) at $u=0$.\\
(a) Real (solid) and imaginary (dashed) parts of the function $\Delta \varepsilon(\beta|u)$ defined by \eqref{deep} at $u=10^{-8}$.
 Vertical lines are located at $\pm \ln(2/u)$.\\
 (b) Real (solid) and imaginary (dashed) parts of the function $\Delta \varepsilon_k(\beta)=\varepsilon_k(\beta)-\exp(\beta)$.
}
\label{kohler}
\end{figure}

Besides the solution $ \varepsilon(\beta|u)$ of equations \eqref{tB}, it is also useful to consider the difference
\begin{equation}\label{deep}
\Delta \varepsilon(\beta|u)=\varepsilon(\beta|u)-u \cosh \beta.
\end{equation}
Figure  \ref{sol} displays its  real and imaginary parts  plotted agains $\beta$ at the small value of the scaling parameter, $u=10^{-8}$.
As one can see from this Figure,  the 
function $\Delta \varepsilon(\beta|u)$ vanishes outside two regions  located  near the points ${\beta\approx\pm R(u)}$, where $R(u)=\ln(2/u)$. 
These two regions become well separated from one another at very small $u$. By this reason, it is useful to shift the argument in the 
solutions of \eqref{tB} by $R(u)$, and to introduce new functions 
\begin{equation}\label{epsR}
\varepsilon_{R}(\beta|u)=\varepsilon\left(\beta+R(u)|u\right), \;\bar{\varepsilon}_{R}(\beta|u)=\bar{\varepsilon}\left(\beta+R(u)|u\right).
\end{equation}
These two functions solve the integral equations, which are obtained from \eqref{tB} by  replacement of the driving term
\begin{equation}\label{uR}
u \cosh \beta \to e^\beta+\frac{u^2}{4}e^{-\beta}
\end{equation}
in their right-hand sides. 

Proceeding in \eqref{epsR} and \eqref{uR} to the limit $u\to0$, one obtains the pseudoenergies
\begin{equation}\label{eps}
\varepsilon_{k}(\beta)=\lim_{u\to0}\varepsilon\left(\beta+R(u)|u\right), \;\bar{\varepsilon}_{k}(\beta)=\lim_{u\to0}\bar{\varepsilon}\left(\beta+R(u)|u\right),
\end{equation}
which correspond to the isotropic point of the XXZ spin chain and solve the TBA integral equations,
\begin{subequations}\label{tBk}
\begin{align}
& \varepsilon_k(\beta)=e^\beta\nonumber
- 2 \int_{-\infty}^{\infty}
d\beta' \, \mathcal{Q}(\beta-\beta') \,
{\rm Re}\,\chi_k(\beta')
 -\\
 &\int_{-\infty}^{\infty}d\beta' \, g(\beta-\beta' - i 0)  \overline{\chi_k(\beta')},\label{ek1}\\
 & \Bar{\varepsilon}_k(\beta)=e^\beta \nonumber
- 2 \int_{-\infty}^{\infty}
d\beta' \, \mathcal{Q}(\beta-\beta')\, {\rm Re}\,\chi_k(\beta')
 -\\
 &\int_{-\infty}^{\infty}d\beta' \, \overline{{g}(\beta-\beta' - i 0) } \chi_k(\beta'),
\end{align}
\end{subequations}
where $\chi_k(\beta)=\ln\left[1+e^{-{\varepsilon}_k(\beta)}\right]$.
After subtraction of $e^\beta$ from  $\varepsilon_k(\beta)$, the resulting difference 
\begin{equation}\label{di}
\Delta \varepsilon_k(\beta)=\varepsilon(\beta)-e^\beta
\end{equation}
decays at $\beta\to\pm \infty$. The real and imaginary parts of this function are plotted in Figure \ref{solk}.
Since the kernels $\mathcal{Q}(\Lambda)$, $g(\Lambda)$ decay  $\sim \Lambda^{-2}$  at $|\Lambda|\to\ \infty$,
the difference \eqref{di} also vanishes as $\Delta \varepsilon_k(\beta)\sim \beta^{-2}$ at large $|\beta|$.  Of course,
the function $ \varepsilon_k(\beta)$ decays $\sim \beta^{-2}$ as well at 
$\beta\to - \infty$, due to \eqref{di}.

As it was mentioned above, the value $Y(0)=-\pi/6$  of the Casimir scaling function at the isotropic point is determined by the CFT.
It is well known \cite{Al_Z90,Klum91,De94}, that the CFT predicted value of finite-size correction to the 
ground-state energy can be alternatively obtained in the TBA approach without explicit solution of the 
integral TBA equations. Let us describe such an alternative derivation of the CFT result  \eqref{pi6}.

The integral representation \eqref{Ysc} of the Casimir scaling function reduces at $u=0$ to the form
\begin{equation}
Y(0) =
 \frac{2}{\pi}\, {\rm Re} \int_{-\infty}^\infty {d\beta}\,e^\beta\, \chi_k'(\beta),\label{Ysc0}
\end{equation}
Let us now rewrite equation \eqref{ek1} as
\begin{align}\nonumber
&e^\beta= \varepsilon_k(\beta)
+ 2 \int_{-\infty}^{\infty}
d\beta' \, \mathcal{Q}(\beta-\beta') \,
{\rm Re}\, \chi_k(\beta')+\\
&\int_{-\infty}^{\infty}d\beta' \,{g}(\beta-\beta' - i 0) \overline{\chi_k(\beta')}.\label{ebet}
\end{align}
The key step of this calculation is the counterintuitive substitution of the
right-hand side  of \eqref{ebet} instead of $e^\beta$ into the integrand in \eqref{Ysc0}. As the result, one obtains
\begin{align*}
&Y(0)=- \frac{2}{\pi}\, {\rm Re}\, \int_{-\infty}^\infty {d\beta}\,\varepsilon_k(\beta)
\, \frac{\varepsilon_k'(\beta)}{1+e^{\varepsilon_k(\beta)}}+\\
&\frac{4}{\pi}\, \iint_{-\infty}^\infty {d\beta}d\beta'\,{\rm Re}\,[\chi_k'(\beta)]\mathcal{Q}(\beta-\beta')\, {\rm Re}\,[\chi_k(\beta')]+\\
&\frac{2}{\pi}\,{\rm Re}\, \iint_{-\infty}^\infty {d\beta}d\beta'\,\chi_k'(\beta)\,g(\beta-\beta'-i0)\, \overline{\chi_k(\beta')}.
\end{align*}
Both double-integrals in the right-hand side vanish due to the kernel symmetry \eqref{symk}.
 After the change of the integration variable $x=\varepsilon_k(\beta)$ in the first integral, we arrive  at the 
 desired CFT result \eqref{pi6},
 \[
Y(0)=- \frac{2}{\pi} \int_{0}^\infty dx\,\frac{x}{1+e^x}=-\frac{\pi}{6}.
\]

For the subsequent analysis, we need the explicit form of the asymptotic expansion of the function $\varepsilon_k(\beta)$ at $\beta\to - \infty$.
We obtained three terms in this expansion by the straightforward perturbative solution of the integral equations \eqref{tBk} at large negative $\beta$. The result reads
\begin{equation}\label{ek}
\varepsilon_k(\beta)=\frac{b_2}{\beta^2}+\frac{b_3}{\beta^3}-\frac{b_2\ln(-\beta)}{\beta^3}+O\left(\beta^{-4}\ln(-\beta)\right),
\end{equation}
where 
\begin{subequations}\label{bb}
\begin{align}
&b_2=-\left(\frac{\pi\, \ln 2}{2}+{\rm Im}\, A_1\right)i,\\
&b_3=\pi\ {\rm Im}\, b_2+i \left(-\pi\, {\rm Re}\, A_1+\frac{ {\rm Im}\, b_2}{2} -{\rm Im}\, A_2\right).
\end{align}
\end{subequations}
Here $A_1$, $A_2$ denote the following converging integrals:
\begin{align}\label{A12}
&A_1=\int_{-\infty}^\infty d\beta \left[
\ln\left(1+e^{-\varepsilon_k(\beta)}\right)-\theta(-\beta)\,\ln 2\right],\\\label{A2}
&A_2=\int_{-\infty}^\infty d\beta\,  \Big[2\beta\,\ln\left(1+e^{-\varepsilon_k(\beta)}\right)-\\\nonumber
&2 \beta \,\theta(-\beta)\,\ln 2+\frac{b_2}{\beta} \theta(-\beta-1)\Big],\nonumber
\end{align}
and $\theta(x)$ is the unit-step function. 

By numerical calculation of the integrals \eqref{A12} and  \eqref{A2} we obtained the  following 
values:
\begin{align}\label{rA1}
&{\rm Re}\, A_1=-0.38806..., \\
&{\rm Im}\, A_1= 0.48200..., \\
&{\rm Im}\, A_2=1.026....\label{iA2}
\end{align}
It turns out, that the   numerical value of 
 ${\rm Im}\, A_1$ is very close to $(\pi/2)(1-\ln 2)=0.482003...$
 \begin{footnote}
 {This fact was  earlier noticed 
 by Kl{\"u}mper  \cite{Klum98} who had calculated the same integral numerically in order to determine
 the low-temperature behaviour of the specific heat in the XXX spin chain.
 The integral $I$ defined by equation (24) in \cite{Klum98} coincides with 
${\rm Im}\, A_1$.}
 \end{footnote}.
 In fact, there are strong arguments \cite{Klum98}  that the latter number is the exact 
 value of the imaginary part of the integral \eqref{A12}, 
\begin{equation}
{\rm Im}\, A_1=\frac{\pi}{2}(1-\ln 2)=0.4820032816....
\end{equation}
Under this assumption, one finds from \eqref{bb}, \eqref{rA1} and \eqref{iA2}
\begin{equation}\label{bb}
b_2=-\frac{\pi i}{2}, \quad b_3 =- \frac{\pi^2}{2}- i\, 0.592....
\end{equation}

We are now ready to return to  the perturbative calculation of the Casimir scaling functions 
$Y(u)$ at a small  $u>0$. First, we write the solution of equations \eqref{tB}
 in the form 
\begin{equation}\label{vps}
  \varepsilon(\beta|u)=\varepsilon^{(0)}(\beta|u)+v(\beta|u), 
\end{equation} 
where
\begin{equation}\label{psen}
\varepsilon^{(0)}(\beta|u)=\varepsilon_k(\beta-R(u))+\overline{\varepsilon_k(-\beta-R(u))},
\end{equation}
is the zero-order term and $v(\beta|u)$ is the small correction. 
The latter could be in principal determined  by means of the perturbative solution of the nonlinear integral equations \eqref{tB},
with the small parameter  $\delta=[R(u)]^{-1}$. 
Next, one could substitute \eqref{vps} into \eqref{Ysc} 
and try to extract several initial terms in the small-$u$ asymptotic expansion for $Y(u)$ from the resulting integrals. 
It turns out, however, that such  direct
perturbative calculations are extremely difficult and not suitable for evaluation of the higher
terms in \eqref{Y0}.  Really, in order to calculate the smallest term $a_4 \delta^{4}$ in  expansion \eqref{Y0}  in this approach, one has to solve perturbatively  the nonlinear integral equations \eqref{tB} to the  fourth  order  in the small parameter $\delta$.

To avoid this problem, we have applied following \cite{De94,Klum98}  the improved  technique, 
which allowed us to obtain  \eqref{Y0} without solving perturbatively  the integral equations \eqref{tB}.  
First,  we rewrite the integral representation  \eqref{Ysc} of the scaling function in the equivalent form, 
\begin{equation}
Y(u) =-\frac{2}{\pi}\, {\rm Re} \int_{-\infty}^\infty {d\beta}\,e^\beta\, \chi_R(\beta|u),\label{Ysc1}
\end{equation}
where $\chi_R(\beta|u)=\ln\left[1+e^{-{\varepsilon}_R(\beta|u)}\right]$.
Then we split the integral $ \int_{-\infty}^\infty {d\beta}$ on the right-hand side into two parts, 
$ \int_{-\infty}^\infty {d\beta}=\int_{-\infty}^{-R(u)} {d\beta}+\int_{-R(u)}^\infty {d\beta}$.  The first 
term is small $\sim u$ due to the [small at $\beta<-R(u)$] factor $e^\beta$ in the integrand in \eqref{Ysc1}.
After integration by parts in the second term, one obtains at $u\to 0$,
\begin{equation}\label{Yu}
Y(u) =\frac{2}{\pi}\, {\rm Re} \int_{-R(u)}^\infty {d\beta}\,e^\beta\, \partial_\beta[\chi_R(\beta|u)]+O(u).
\end{equation}
This formula extends  \eqref{Ysc0} to the case of a small positive $u$.  
Recall next, that the function ${\varepsilon}_R(\beta|u)$ defined by \eqref{epsR} solves the integral equation \eqref{tB1}
modified according to  \eqref{uR}. Let us rewrite this equation in the  form
similar to \eqref{ebet},
\begin{align}
&e^\beta= \varepsilon_R(\beta|u)-\frac{u^2}{4}e^{-\beta}
+ \\\nonumber
& \int_{-\infty}^{\infty}
d\beta' \,\Big[ 2\mathcal{Q}(\beta-\beta') \,
{\rm Re}\, \chi_R(\beta'|u)+\\
&{g}(\beta-\beta' - i 0) \,\overline{\chi_R(\beta'|u)}\Big].\nonumber
\end{align}
After substitution of the right-hand side instead of $e^\beta$ in the integrand in \eqref{Yu} and
straightforward calculations, one finds
\begin{equation}
Y(u) =-\frac{\pi}{6}+\Delta Y(u)+O([R(u)]^{-5}),
\end{equation}
where
\begin{align}\label{Yu1}
&\Delta Y(u) =\frac{2}{\pi}\,{\rm Re} \int_{0}^\infty {d\beta}\!\int_{-\infty}^\infty d\beta'\,
\Big\{2\mathcal{Q}(\beta-\beta')\partial_\beta \chi(\beta|u)\cdot\\ \nonumber
&\,{\rm Re}\,[\chi(\beta'|u)]+
g(\beta-\beta'-i0)\,
\partial_\beta[\chi(\beta|u)]\, \overline{\chi(\beta'|u)}\Big\},\nonumber
\end{align}
and $\chi(\beta|u)$ is given by \eqref{ch}.
In contrast to the $u=0$ case, the  double integral in the right-hand side of \eqref{Yu1} is nonzero at $u>0$ due to the finite  limits of integration.   
However, this integral  decreases with decreasing $u$ and vanishes  at $u=0$. 

Using integration by parts and the symmetry properties \eqref{symk} and \eqref{syme}, the double-integral  in 
\eqref{Yu1} can be conveniently represented as the sum of three terms,
\begin{equation}\label{dYA}
\Delta Y(u)=A+B+C,
\end{equation}
where
\begin{subequations}\label{ABC}
\begin{align}
&A=\frac{2}{\pi }[\chi(0|u)]^2\,{\rm Re} \int_{2 R(u)}^\infty d\beta\, U(\beta),\\
&B=-\frac{2}{\pi }\chi(0|u)\,{\rm Re} \int_0^\infty d\beta\, U(\beta+R(u)) \Psi(\beta|u),\\
&C=\frac{4}{\pi }\iint_0^\infty d\beta d\beta'\, \mathcal{Q}(\beta+\beta') {\rm Re}\, [\partial_\beta\chi(\beta|u)]{\rm Re}\, [\Psi(\beta'|u)]+\nonumber\\
&\frac{2}{\pi }{\rm Re}\iint_0^\infty d\beta d\beta' g(\beta+\beta'-i0)\partial_\beta\chi(\beta|u)\Psi(\beta'|u),
\end{align}
\end{subequations}
and 
\begin{align}
&U(\beta)=2 \mathcal{Q}(\beta)+g(\beta),\\
&\Psi(\beta|u)=\chi(\beta|u)-\theta\big(R(u)-\beta\big)\,\chi(0|u).
\end{align}
We substituted the function $\varepsilon(\beta|u)$ in the form \eqref{vps} into 
the integrals in \eqref{ABC} and expanded the results in the small parameter $\delta=[R(u)]^{-1}$ to the fourth order.
It turns out that, up to this order, (i) the correction term $v(\beta|u)$ in \eqref{vps} does not contribute to 
these integrals and (ii) these integrals can be expressed solely in terms of two numbers ${\rm Im}\, b_2$ and 
${\rm Im}\, b_3$, which characterise the asymptotical behaviour of the function $\varepsilon_k(\beta)$
at $\beta\to-\infty$; see \eqref{ek} and \eqref{bb}. As the result, we obtained,
\begin{align}
&A=\frac{\pi (\ln 2)^2\,\delta^3}{16}+O(\delta^{5}),\\
&B=\ln 2 \,\bigg\{-
\frac{(2\,{\rm Im}\, b_2+\pi \ln 2)\,\delta^3}{16}+\\
&\frac{[-12\, {\rm Im}\, b_2 \ln(2/\delta)+ 3\,{\rm Im}\, b_2+12\, {\rm Im}\, b_3 ]\,\delta^4}{128} \nonumber
\bigg\}
+o(\delta^{4}),\nonumber\\
&C=\frac{\delta^3\, {\rm Im}\, b_2\,[2 \,{\rm Im}\, b_2+\pi \ln 2]}{8\pi}-\\
&\frac{\delta^4}{8\pi}\bigg\{
-3\,  {\rm Im}\, b_2\, \ln(2/\delta)\left({\rm Im}\, b_2+\frac{\pi\, \ln 2}{4}
\right)+\nonumber\\
&\frac{\pi \ln 2}{16}\left(
5\, {\rm Im}\, b_2+12\,{\rm Im}\, b_3
\right)+\nonumber\\
&{\rm Im}\, b_2 ({\rm Im}\, b_2+3\, {\rm Im}\, b_3)
\bigg\}
+o(\delta^{4})\nonumber.
\end{align}
This yields for \eqref{dYA},
\begin{align}
&\Delta Y(u)=\frac{({\rm Im}\, b_2)^2 \delta^3}{4\pi}+\frac{3\,({\rm Im}\, b_2)^2\, \delta^4 \ln(2/\delta)}{8\pi}-\\
&\frac{ \delta^4\,{\rm Im}\, b_2}{64 \pi}\,\left (
8\, {\rm Im}\, b_2+24 \, {\rm Im}\, b_3+\pi \ln 2
\right)+o(\delta^4).  \nonumber
\end{align}
After substitution of the obtained earlier values \eqref{bb} 
of the constants ${\rm Im}\, b_2,\, {\rm Im}\, b_3$ into this result, we arrive finally at \eqref{Y0}.

To conclude this section, we comment on the perturbative analysis around the CFT critical point 
of two similar TBA equations, which were previously performed
 by Kl\"umper \cite{Klum98} and by Destri and de Vega \cite{De94}.
 
The nonlinear integral TBA equation (3) in  \cite{Klum98} studied by Kl\"umper describes the 
thermodynamic properties of  the infinite
 antiferromagnetic isotropic spin-1/2 Heisenberg chain in the presence of a uniform magnetic field.
In the case of zero magnetic field, this equation differs from equation \eqref{tBA2} only by one term. Namely,  
the  driving term $u \cosh \beta$ in  the right-side of \eqref{tBA2} replaces \begin{footnote}
{Kl{\"u}mper uses in  \cite{Klum98} the notation $x$ for the rapidity variable ($\beta$ in our notations), and notations 
$\frak{a}(x)$, $\frak{A}(x)$ 
for the functions, which are analogous to ours $e^{-\bar{\varepsilon}(\beta)}$, and $\overline{\chi (\beta)}$, respectively. }
\end{footnote}  the term  $\frac{\pi J/T }{\cosh \beta}$  in equation (3) in  \cite{Klum98}. Despite this difference,  the low-temperature asymptotical analysis of the nonlinear TBA equation (3)  presented  in Section 2  of \cite{Klum98} has some similarities with 
our small-$u$ perturbative calculations described in this Section. In particular, the small parameters
$\frac{T}{\pi \,J}$ and $\mathfrak{L}^{-1}=1/\ln(\pi J/T)$ used in Section~2 of \cite{Klum98} are analogous to
the small parameters $u$ and $\delta=1/\ln(2/u)$, which we have exploited in the described above calculations. 
Note finally, that we have calculated four temperature-dependent terms in the asymptotic expansion \eqref{def}
for the free energy, while only two such terms were obtained in the analogous expansion (26) in 
\cite{Klum98}.

The nonlinear integral TBA equation for the sine-Gordon (massive Thiring) model was obtained by  Destri  and  de Vega, see equation  (5.12) in  \cite{De94}. Its asymptotical analysis 
close to  the conformal regime was performed by these 
authors in Section 7.3. While the  driving term  in this equation is the same as in
ours equations (\ref{tBa}), the  kernels $G_0(\Lambda,\gamma)$,  $G_1(\Lambda,\gamma)$  
(see equation (5.13) in \cite{De94}  and the non-numbered foregoing equation there \begin{footnote}
{ Note that the latter equation in  \cite{De94} contains misprints, which are corrected below 
in the second line of equation (\ref{G0})}\end{footnote})  in the integral terms are different. Namely, 
\begin{eqnarray}\label{G0}
&&G_0(\Lambda,\gamma)=\frac{1}{2\pi i} \frac{\partial \ln S(\Lambda,\gamma_s(\gamma))}{\partial\Lambda}=\\
&&\int_{-\infty}^\infty \frac{dk}{4\pi}\, \frac{\sinh\left[\left(\frac{\pi^2}{2\gamma}-\pi\right) k\right]}
{\sinh\left[\left(\frac{\pi^2}{2\gamma}-\frac{\pi}{2}\right) k\right]\, \cosh(\pi k/2)} \,e^{i k \Lambda}, \nonumber\\
&& G_1(\Lambda,\gamma)=G_0(\Lambda+i\pi,\gamma),
\end{eqnarray}
where $S(\Lambda,\gamma_s)$ is the soliton-soliton scattering amplitude \eqref{Sc} in the sine-Gordon model,
and parameters $\gamma$ and $\gamma_s$ are related according to \eqref{gam}. In the limit $\gamma\to0$, the
integral kernel \eqref{G0} degenerates to the form \eqref{ps},
\[
\lim_{\gamma\to0} G_0(\Lambda,\gamma) =\mathcal{Q}(\Lambda).
\]
So, the integral nonlinear  TBA equations \eqref{tBa} describing the scaling behaviour of the 
gapped XXZ spin chain represent the degenerate $\gamma\to0$ limiting case of the TBA equations
for the sine-Gordon model derived by Destri and de Vega. 
However, our small-$u$ asymptotical analysis described in this section is to a large  extent different from that
 developed by Destri and de Vega in Section 7.3  in  \cite{De94}. The reason is that  
in our case the integral kernel $\mathcal{Q}(\Lambda)=G_0(\Lambda,0)$  decays   slowly $\sim \Lambda^{-2}$ at large 
rapidities $|\Lambda|$, 
whereas in the non-degenerate case  $\gamma>0$ studied in  \cite{De94}, 
the kernel \eqref{G0} exponentially vanishes at 
 $|\Lambda|\to\infty$; see the non-numbered equations between (7.24) and 
(7.25) in  \cite{De94}. 
\section{ \label{RenGr} Field-theortical derivation of \eqref{EA}}
The continuous limit of the XXZ spin chain \eqref{XXZH} near the isotropic point 
$\Delta=1$ can be described 
by the  marginal perturbation of the Gaussian CFT \cite{Aff98,Luk98},
\begin{equation}\label{Hs}
{H}={H}_{WZW}-\frac{8\pi^2}{\sqrt{3}}\int_0^{L_x} dx\,[g_x (J^x \bar{J}^x+ J^y \bar{J}^y)+g_z J^z \bar{J}^z]. 
\end{equation}
Here $H_{WZW}$ 
is the Hamiltonian of free bosons compactificated at the radius $\mathcal{R}=1/{\sqrt{2\pi}}$, 
 or equivalently, the $SU(2)$  Wess-Zumino-Witten (WZW) Hamiltonian of level
$k=1$.  Operators  $J^\mathfrak{a}$ and $\bar{J}^\mathfrak{a}$,  with $\mathfrak{a}=x,y,z$, 
represent the components of the holomorphic and anti-holomorphic currents, respectively. 
Their normalization can be fixed by the Operator Product Expansions (OPE),
\begin{align}\label{OPE}
J^\mathfrak{a}(z)J^{\mathfrak b}(z')=
\frac{\delta^{{\mathfrak a}{\mathfrak b}}}{8\pi^2 (z-z')^2}+\frac{i \epsilon^{{\mathfrak a}{\mathfrak b}
{\mathfrak c}}}{2\pi (z-z')}\,J^{\mathfrak c}(z)+\ldots,\\
\bar{J}^\mathfrak{a}(\bar{z})\bar{J}^{\mathfrak b}(\bar{z}')=
\frac{\delta^{{\mathfrak a}{\mathfrak b}}}{8\pi^2 (\bar{z}-\bar{z}')^2}+\frac{i \epsilon^{a{\mathfrak b}{\mathfrak c}}}{2\pi (\bar{z}-\bar{z}')}\,\bar{J}^{\mathfrak c}(\bar{z})+\ldots.\nonumber
\end{align}
Following  \cite{Affleck_1989,Aff98}, the  normalization in equation (\ref{Hs}) has been chosen to ensure  that the operators multiplying $g$ in the isotropic case (see equation \eqref{HXXX} below) have a correlation function with unit amplitude. 
\begin{figure}[htb]
\centering
\includegraphics[width=\linewidth, angle=00]{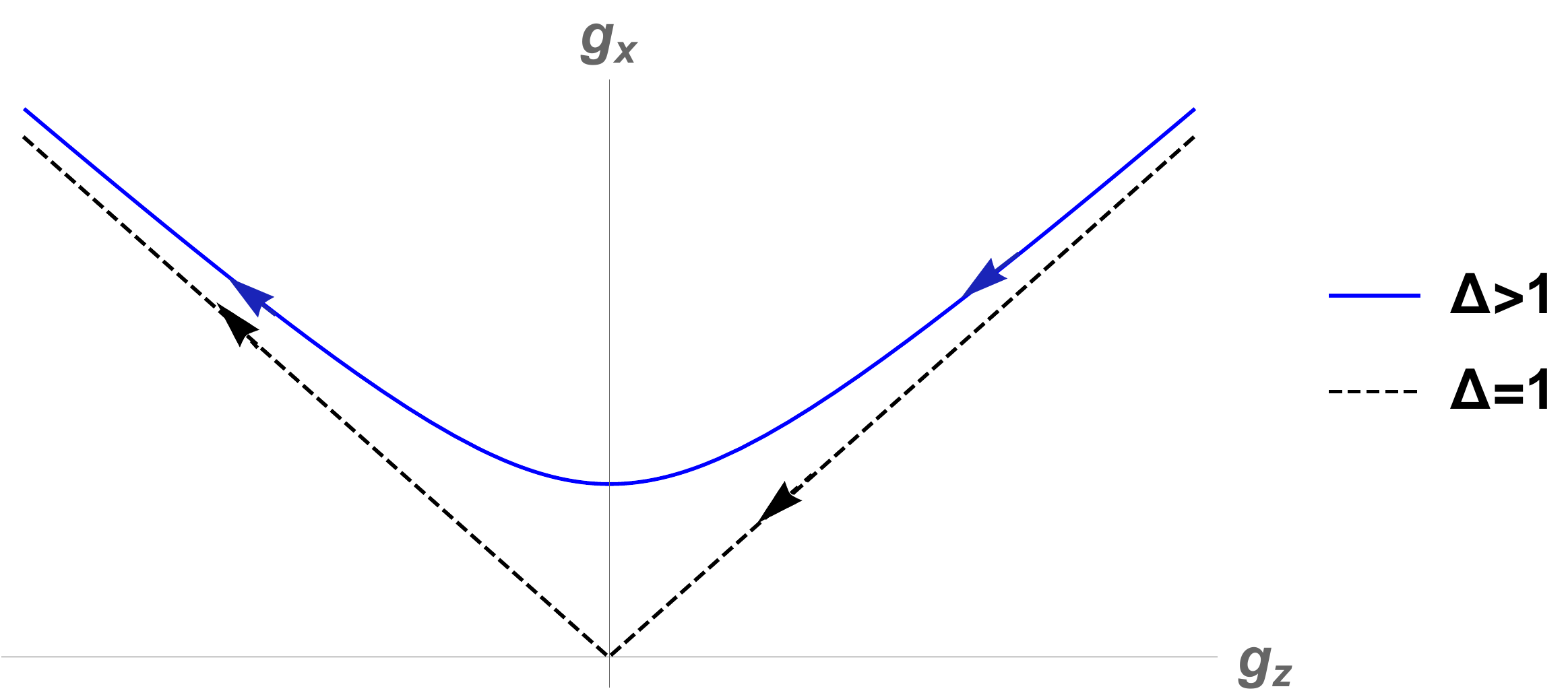}
\caption{\label{fig:RG} Kosterlitz-Thouless RG flow  corresponding to equations \eqref{RGa}
 at $\Delta>1$ and $\Delta=1$.} 
\end{figure}
The Renormalization Group (RG) flow 
of the scaling parameters in \eqref{Hs} in the one-loop approximation is described by the Kosterlitz-Thoulless RG equations \cite{Aff98}, 
 \begin{eqnarray} \label{RGa}
 &&\beta_x\equiv d\, g_x/dr =-\frac{4\pi}{\sqrt{3}} \,g_x\, g_z,\\\nonumber
 &&  \beta_z\equiv d\, g_z/dr =-\frac{4\pi}{\sqrt{3}}\, g_x^2,
 \end{eqnarray}
 where $r=\ln {L}$, and ${L}$ is the length scale.
 
 Two RG trajectories are shown in Figure \ref{fig:RG}. The dashed bisector of the first quadrant  $g_x=g_z\equiv g>0$   corresponds to the isotropic point $\Delta=1$ of the spin-chain Hamiltonian \eqref{XXZH}. 
The RG equations \eqref{RGa}  reduce in the isotropic case to the simple equation,
\begin{equation} \label{RG2}
\frac{dg(r)}{dr }=-\pi b \,g^2(r),
\end{equation}
with  $b=4/\sqrt{3}$.
Its solution taking the value $g(r_0)$ at the initial point $r_0\simeq \ln a$ reads as,
\begin{equation}
g(r)=\frac{g(r_0)}{1+\pi \,b \,g(r_0)(r-r_0)}.
\end{equation}
The leading asymptotics of this solution at  $r\to\infty$ does not depend on $g(r_0)$, 
\begin{equation}\label{gimp}
 g(r)=\frac{1}{\pi b\, (r-r_0)}+O(r^{-2}).
\end{equation}
Along the critical line, the effective Hamiltonian 
\eqref{Hs} reduces to the form
\begin{equation}\label{HXXX}
{H}_{XXX}={H}_{WZW}-g\int_0^{L_x} dx\,\varphi(x),
\end{equation}
where $\varphi(x)$ is the  marginally irrelevant operator,
\begin{equation}
\varphi(x)=\frac{8\pi^2}{\sqrt{3}} \,\mathbf{J}(x)\cdot\bar{\mathbf{J}}(x).
\end{equation}
 In the Euclidean plane, its two- and three-point correlation functions are fixed due to \eqref{OPE},
\begin{align}
&\langle \varphi({\mathbf r}_1)\varphi({\mathbf r}_2)\rangle=\frac{1}{|{\mathbf r}_{1,2}|^4}, \\
&\langle \varphi({\mathbf r}_1)\varphi({\mathbf r}_2)\varphi({\mathbf r}_3)\rangle=-
\frac{b}{|{\mathbf r}_{1,2}|^2|{\mathbf r}_{1,3}|^2|{\mathbf r}_{2,3}|^2},
\end{align}
where ${\mathbf r}_{i,j}={\mathbf r}_{i}-{\mathbf r}_{j}$.

 Affleck {\it et al.} \cite{Affleck_1989} considered the generalisation of the effective Hamiltonian \eqref{HXXX} to the case of the WZW model with arbitrary positive integer $k$, and  performed for it 
the perturbative  calculation of  the ground-state energy $E_0$   to the third order in $g$. 
In the case $k=1$, their result (see the non-numbered equation between equations  (8) and (9) in 
\cite{Affleck_1989}) reads, 
\begin{equation}\label{E0A}
E_0(g)= e_0 L_x-\frac{\pi}{6 L_x}-\frac{\pi^4 }{3 L_x}\, b g^3+\ldots,
\end{equation}
where $e_0$ denotes the non-universal bulk energy density in the infinite system. 
Note that  Affleck {\it et al.}   \cite{Affleck_1989}  did not present  the details of their calculation of $E_0(g)$.
However, similar perturbative calculations were described earlier by Cardy \cite{Cardy86,Cardy87}, 
and in the most detailed form by Ludwig and Cardy   \cite{Lud87}.  After replacement of $g$ in
\eqref{E0A} by its renormalisation group improved
value $g\to 1/[\pi b \ln (L/a)]$ with $L\sim L_x$ in accordance with \eqref{gimp}, the authors of \cite{Affleck_1989}
arrived finally at \eqref{Ei}. This result was later confirmed by Lukyanov \cite{Luk98}.

Let us  turn now to the anisotropic case $0<\eta\ll1$, and show how the asymptotic formula 
\eqref{EA} can be derived following the strategy outline above. 
To this end, 
 consider the RG flow in the massive antiferromagnetic phase $\Delta>1$,  which is illustrated 
by the upper trajectory  in Figure \ref{fig:RG}. 
The first integral 
$A_x=g_x^2-g_z^2$
of the  Kosterlitz-Thoulless RG equations \eqref{RGa} remains positive along it. 
The solution
of the RG equations  \eqref{RGa}, which corresponds to this trajectory  reads,
\begin{eqnarray}\label{gt}
&&g_z(r)=-\sqrt{A_x}\, \tan\left(4\pi r \sqrt{\frac{A_x}{3}}
\right),  \\
&&g_x(r)=\sqrt{A_x+g_z^2(r)},\nonumber
\end{eqnarray} 
with $r$ varying in the interval $r_{min}<r<r_{max}$, where 
\[
r_{max}=-r_{min}=\frac{1}{8}\sqrt{\frac{3}{A_x}}.
\]
The well-known arguments (see p. 124 in \cite{cardy1996scaling}) lead to the requirement  $r_{max}-r_{min}\simeq\ln[\xi(\eta)/a]$, which 
 allows one together with \eqref{clas} and \eqref{met}  to relate parameters $A_x$ and $\eta$,
\begin{equation}
A_x=\frac{3\eta^2}{4 \pi^4}.
\end{equation} 
Note also that
\begin{equation}
r-r_{min}\simeq\ln[L/a], \quad  r_{max}-r\simeq\ln[\xi(\eta)/L]. \label{rL} 
\end{equation} 

Let us choose now 
the running point $\{g_z(r),g_x(r)\}$ in the upper RG trajectory in such a way,  that: 
\begin{eqnarray}
&& g_z(r)<0, \; {\mathrm {and}}\;\; g_x(r)>0, \\
&& \sqrt{A_x}\ll g_x(r)\ll 1.
\end{eqnarray}
Under these conditions, the RG trajectory approaches its asymptote $g_x=-g_z$ in the second quadrant, 
the argument of the tangent in equation \eqref{gt} lies slightly below 
its pole at $\pi/2$, 
and one finds from  \eqref{gt}-\eqref{rL},
\begin{eqnarray}\label{gz}
&&g_z(r)\cong-\frac{\sqrt{3}}{4\pi (r_{max}-r)}\cong-\frac{\sqrt{3}}{4\pi \ln[\xi(\eta)/L]},\\
&&g_x(r)\cong-g_z(r)\cong \frac{\sqrt{3}}{4\pi \ln[\xi(\eta)/L]},\label{gx}
\end{eqnarray}
where $\xi(\eta)/L\gg 1$. For such a choice of the scaling variables $g_x$, $g_z$, 
we can approximately represent the effective Hamiltonian \eqref{Hs}  in the form
\begin{equation}\label{Hxxz}
H=H_{WZW}-g_x\int_0^{L_x} dx\,\Phi(x),
\end{equation} 
where $g_x>0$, and $\Phi(x)$ is the following marginally relevant operator
 \begin{equation} \label{Phi}
\Phi(x)=\frac{8\pi^2}{\sqrt{3}}\left[J^x(x)\bar{J}^x(x)+J^y(x)\bar{J}^y(x)-J^z(x)\bar{J}^z(x)
 \right]
 \end{equation}
Its two- and three-point correlation functions  in the plane can be easily found from  \eqref{OPE}, 
\begin{eqnarray}
&&\langle \Phi({\mathbf r}_1)\Phi({\mathbf r}_2)\rangle=\frac{1}{3|{\mathbf r}_{1,2}|^4}, \\
&&\langle \Phi({\mathbf r}_1)\Phi({\mathbf r}_2)\Phi({\mathbf r}_3)\rangle=-
\frac{b_\Phi}{|{\mathbf r}_{1,2}|^2|{\mathbf r}_{1,3}|^2|{\mathbf r}_{2,3}|^2},
\end{eqnarray}
where $b_\Phi=-4/\sqrt{3}$. Note that equation \eqref{gx}
can be rewritten as, 
\begin{equation}
g_x(r)\cong \frac{1}{\pi\, b_\Phi \ln[L/\xi(\eta)]}.\label{gx1}
 \end{equation}
 
 The ground-state energy of the Hamiltonian \eqref{Hxxz} can be calculated perturbatively in 
 the small parameter $g_x$. This   calculation literally reproduces the derivation of equation 
 \eqref{E0A} outlined above, in which one should replace  $\varphi(x)\to \Phi(x)$,
 $g\to g_x$, and $ b\to b_\Phi$. Accordingly, one obtains instead of  \eqref{E0A},
 \begin{equation}\label{EAf}
 E_0^{(af)}(L_x,g_x)= e_0\,L_x-\frac{\pi}{6 L_x}-\frac{\pi^4}{3 L_x}\, b_\Phi\, g_x^3+\ldots.
 \end{equation} 
 After  further replacement of the scaling variable $g_x$  by its  RG improved value \eqref{gx1}
 and setting $L\sim L_x$, 
 one arrives at the final result \eqref{EA}.

\end{document}